\documentclass[12pt]{iopart}
\usepackage{iopams}
\usepackage{calc}
\usepackage{graphicx}
\usepackage{bm}

\newcommand{\figuresize}{\columnwidth}

\newcommand{\CGT}{Cr$_{2}$Ge$_{2}$Te$_{6}$}
\newcommand{\CST}{Cr$_{2}$Si$_{2}$Te$_{6}$}

\newcommand{\BT}{Bi$_{2}$Te$_{3}$}
\newcommand{\wavnum}{cm$^{-1}$}

\newcommand{\methodsec}{Method Section}

\newcommand{\splittemperature}{near \Tc}

\newcommand{\CCO}{CdCr$_{2}$O$_{4}$}
\newcommand{\ZCO}{ZnCr$_{2}$O$_{4}$}
\newcommand{\ZCS}{ZnCr$_{2}$S$_{4}$}
\newcommand{\Tc}{T$_{C}$}

\newcommand{\spinphononcoupling}{H_{int} = \sum\limits_{i,\delta}\frac{\partial J}{\partial u}(\mathbf{S}_{i}^{a}{}\cdot\mathbf{S}_{i+\delta}^{b})u\label{equ:hamitionian_sp}
}

\begin{document}

\title{Magneto-Elastic Coupling in a potential ferromagnetic 2D Atomic Crystal} 

\author{Yao Tian}
\address{Department of Physics, University of Toronto, ON M5S 1A7 Canada}

\author{Mason J. Gray}
\address{Department of Physics, Boston College 140 Commonwealth Ave Chestnut Hill MA 02467-3804 USA}

\author{Huiwen Ji}
\address{Department of Chemistry, Princeton University, Princeton, NJ 08540 USA}

\author{R. J. Cava}
\address{Department of Chemistry, Princeton University, Princeton, NJ 08540 USA}



\author{Kenneth S. Burch}
\address{Department of Physics, Boston College 140 Commonwealth Ave Chestnut Hill MA 02467-3804 USA}
\ead{ks.burch@bc.edu}


\begin{abstract}

\CGT{} has been of interest for decades, as it is one of only a few naturally forming ferromagnetic semiconductors. Recently, this material has been revisited due to its potential as a 2 dimensional semiconducting ferromagnet and a substrate to induce anomalous quantum Hall states in topological insulators.
However, many relevant properties of \CGT{} still remain poorly understood, especially the spin-phonon coupling crucial to spintronic, multiferrioc, thermal conductivity, magnetic proximity and the establishment of long range order on the nanoscale. We explore the interplay between the lattice and magnetism through high resolution micro-Raman scattering measurements over the temperature range from 10 K to 325 K. Strong spin-phonon coupling effects are confirmed from multiple aspects: two low energy modes splits in the ferromagnetic phase,  magnetic quasielastic scattering in paramagnetic phase, the phonon energies of three modes show clear upturn below \Tc{}, and the phonon linewidths change dramatically below \Tc{} as well.  Our results provide the first demonstration of spin-phonon coupling in a potential 2 dimensional atomic crystal.
\end{abstract}

\section{\label{sec:intro}Introduction}
\CGT{} is a particularly interesting material since it is in the very rare class of ferromagnetic semiconductors and possesses a layered, nearly two dimensional structure due to van Der Waals bonds\cite{CGT_original,li2014crxte}. Recently this material  has been revisited as a substrate for the growth of the topological insulator \BT{} to study the anomalous quantum Hall effect\cite{BT_CGT_quantum_hall}. Furthermore the van Der Waals bonds make it a candidate two dimensional atomic crystal, which is predicted as a platform to study 2D semiconducting ferromagnets and for single layered spintronics devices\cite{sivadas2015magnetic}. In such devices, spin-phonon coupling can be a key factor in the magnetic and thermal relaxation processes\cite{golovach2004phonon,ganzhorn2013strong,jaworski2011spin}, while generating other novel effects such as multiferroticity\cite{wesselinowa2012origin,issing2010spin}. Combined with the fact that understanding heat dissipation in nanodevices is crucial, it is important to explore the phonon dynamics and their interplay with the magnetism in \CGT{}.  Indeed,  recent studies have shown the thermal conductivity of its cousin compound \CST{} linearly increases with temperature in the paramagnetic phase, suggesting strong spin-phonon coupling is crucial in these materials\cite{casto2015strong}. However there are currently no direct probes of the phonon dynamics of \CGT{}, let alone the spin-phonon coupling. Such studies are crucial for understanding the potential role of magneto-elastic effects that could be central to the magnetic behavior of \CGT{} as a 2D atomic crystal and potential nano magneto-elastic device.  Polarized temperature dependent Raman scattering is perfectly suited for such studies as it is well established for measuring the phonon dynamics and  the spin-phonon coupling in bulk and 2D atomic crystals\cite{compatible_Heterostructure_raman,Raman_Characterization_Graphene,Raman_graphene,Pandey2013,sandilands2010stability,zhao2011fabrication,calizo2007temperature,sahoo2013temperature,polarized_raman_study_of_BFO,dresselhaus2010characterizing}.  Compared to other techniques, a high resolution Raman microscope can track sub-\wavnum{} changes to uncover subtle underlining physics. A demonstration of Raman studies of \CGT{} can be extremely meaningful for the future study of the exfoliated 2D ferromagnets.

\begin{figure}
    \centering
    \includegraphics[width=0.5\columnwidth]{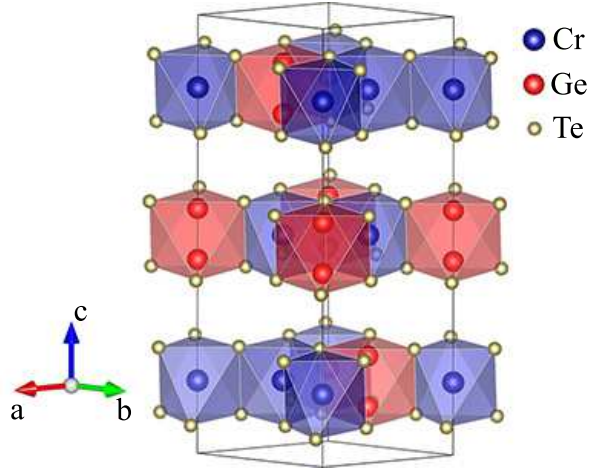}
    \caption{Crystal structure of \CGT{}. The unit cell is indicated by the black frame. Cr and Ge dimer are inside the octahedra formed by Te atoms. One third of the octahedra are filled by Ge-Ge dimers while the other is filled by Cr ions forming a distorted honeycomb lattice.}
    \label{fig:CGT_structure}
\end{figure}

\begin{figure}
  \centering
   \includegraphics[width=\figuresize]{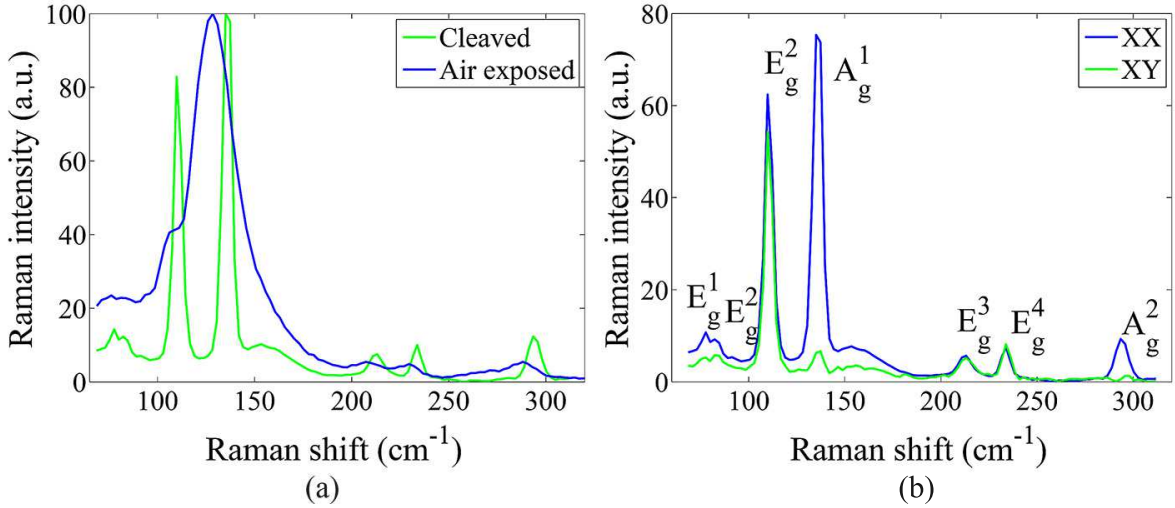}\\
  \caption{(a): Raman spectra of \CGT{} taken in different conditions. All the spectra are taken at 300 K. The Raman spectra of the air-exposed sample shows broader and fewer Raman modes, indicating the formation of oxides. (b): Normalized Raman Spectra of \CGT{} in XX and XY geometry at 270 K, showing the different symmetry of the phonon modes.}\label{633_532_oldsample_raman}
\end{figure}

In this paper, we demonstrate the ease of exfoliating \CGT{}, as well as the dangers of doing so in air. Namely we find the Raman spectra of \CGT{} are strongly deteriorated by exposure to air, but \CGT{} exfoliated in a glovebox reveals bulk like spectra.  In addition, we find strong evidence for spin-phonon coupling in bulk \CGT{}, via  polarized temperature dependent Raman spectroscopy. The spin-phonon coupling has been confirmed in multiple ways: below \Tc{} we observe a split of two phonon modes due to the breaking of time reversal symmetry; a drastic quenching of magnetic quasielastic scattering; an anomalous hardening of an additional three modes; and a dramatic decrease of the phonon lifetimes upon warming into the paramagnetic phase.  Our results also suggest the possibility of probing the magneto-elastic coupling using Raman spectroscopy, opening the door for further studies of exfoliated 2D \CGT{}.


\section{\label{sec:exp}\methodsec}
Single crystal \CGT{} was grown with high purity elements mixed in a molar ratio of 2:6:36; the extra Ge and Te were used as a flux. The materials were heated to 700$^{o}$C for 20 days and then slow cooled to 500$^{o}$C over a period of 1.5 days. Detailed growth procedures  can be found elsewhere\cite{Huiwen_doc}.  The Raman spectra on the crystal were taken in a backscattering configuration with a home-built Raman microscope\cite{RSI_unpublished}. The spectra were recorded with a polarizer in front of the spectrometer. Two Ondax Ultra-narrow-band Notch Filters were used to reject Rayleigh scattering. This also allows us to observe both Stokes and anti-Stokes  Raman shifts and provides a way to confirm the absence of local laser heating.  A solid-state 532 nm laser was used for the excitation. The temperature variance was achieved by using an automatically controlled closed-cycle cryostation designed and manufactured by Montana Instrument, Inc. The temperature stability was within 5 mK. To maximize the collection efficiency, a 100x N.A. 0.9 Zeiss objective was installed inside the cryostation. A heater and a thermometer were installed on the objective to prevent it from being damaged by the cold and to keep the optical response uniform at all sample temperatures.  The laser spot size was 1 micron in diameter and the power was kept fairly low (80 $\mu$W) to avoid laser-induced heating. This was checked at 10 K by monitoring the anti-Stokes signal as the laser power was reduced. Once the anti-Stokes signal disappeared, the power was cut an additional $\approx 50\%$.

\begin{figure}
    \centering
    \includegraphics[width=\textwidth]{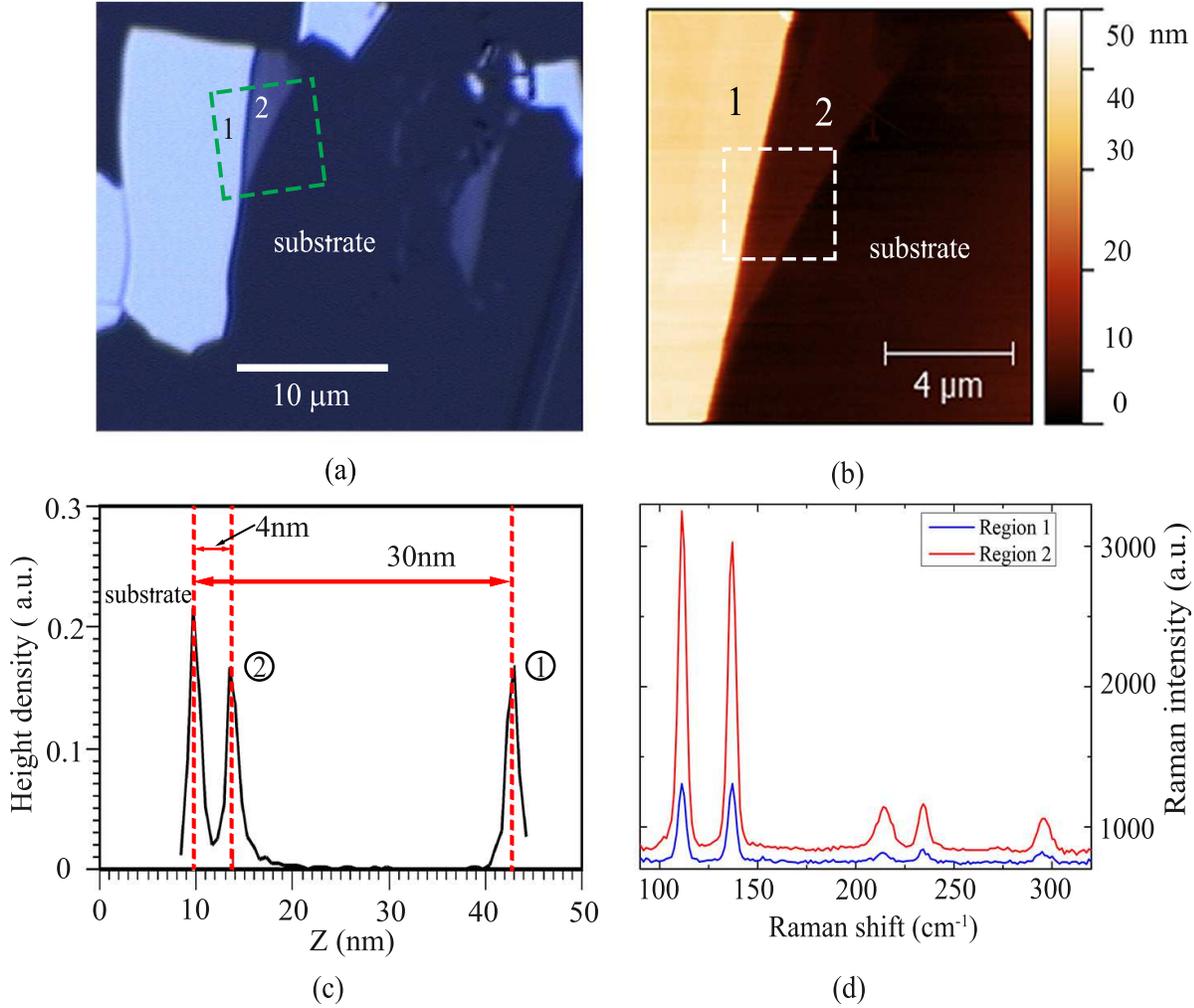}
    \caption{(a): The reflection optical image of the exfoliated \CGT{}. The texts indicate different position of the prepared samples. (b): AFM topography image of the rectangle region in a. (c): The height distribution of the rectangle region in b. The height difference between the peaks reveals the thickness of the \CGT{} flakes, which are region 1: 30 nm and region 2: 4 nm. (d): Raman spectra of the two exfoliated \CGT{} flakes }
    \label{fig:exfoliated_CGT}
\end{figure}

\section{\label{sec:Results_and_discussion}Results}
\subsection{Raman studies at room temperature}

We first delve into the lattice structure of \CGT{} (shown in Fig. \ref{fig:CGT_structure}). This material contains van der Waals bonded layers, with the magnetic ions (Cr, \Tc{}=61 K) forming a distorted honeycomb lattice\cite{Huiwen_doc}. The Cr atoms are locally surrounded by Te octahedra, and thus the exchange between Cr occurs via the Te atoms. Based on the group theory analysis, the Raman-active modes in \CGT{} are of A$_{g}$, E$_{1g}$ and E$_{2g}$ symmetry, and  E$_{1g}$ and  E$_{2g}$ are protected by time-reversal symmetry. In the paramagnetic state we expect to see 10 Raman-active modes, because the E$_{1g}$ and E$_{2g}$ mode are not distinguishable by energy (see details in the supplemental materials).

Keeping the theoretical analysis in mind, we now turn to the mode symmetry assignment of \CGT{}. This analysis was complicated by the oxidation of the \CGT{} surface.  Indeed, many chalcogenide materials suffer from easy oxidation of the surface, which is particularly problematic for \CGT{} as TeO$_{x}$ contains a strong Raman signal\cite{Raman_aging_effect}. The role of oxidation and degradation are becoming increasingly important in many potential 2D atomic crystals\cite{osvath2007graphene}, thus a method to rapidly characterize its presence is crucial for future studies. For this purpose, we measured the Raman response at room temperature in freshly cleaved, as well as air-exposed \CGT{} (shown in Fig. \ref{633_532_oldsample_raman}a). The air-exposed sample reveals fewer phonon modes which are also quite broad, suggesting the formation of an oxide. A similar phenomena was also observed in similar materials and assigned to the formation of  TeO$_{x}$\cite{Raman_amorphous_crystalline_transition_CGTfamily}.  

\begin{figure}[!ht]
  \centering
  \includegraphics[width=\textwidth]{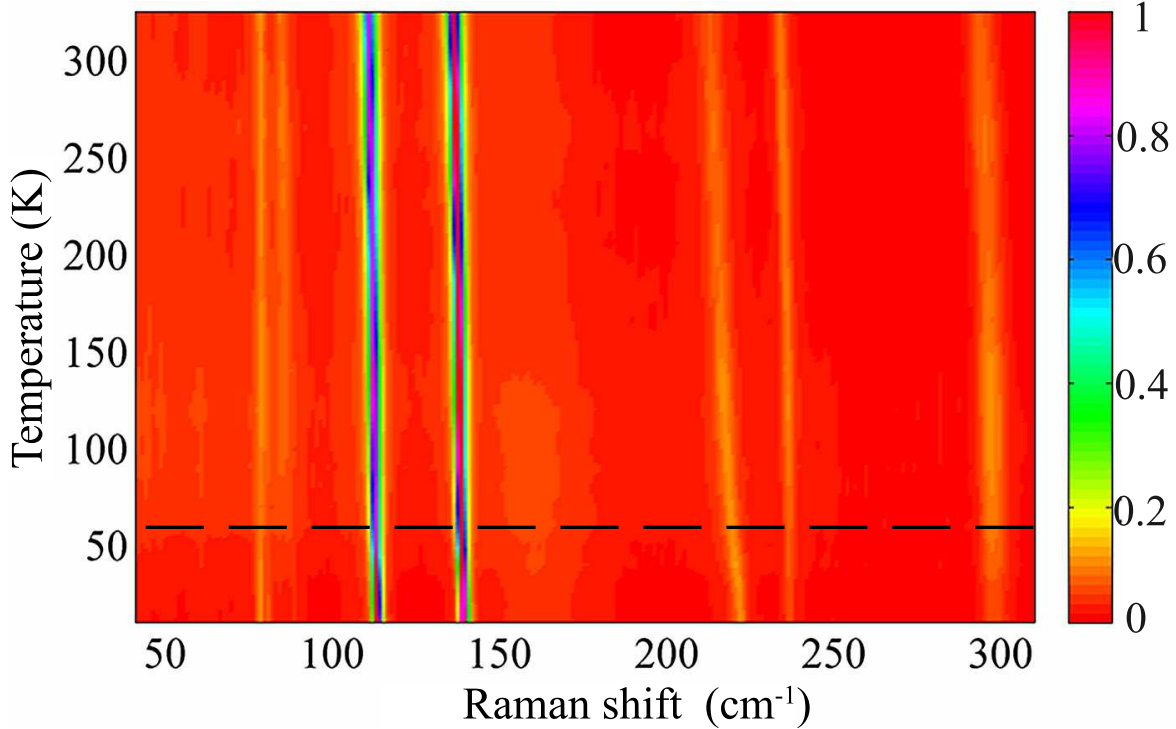}\\
  \caption{Temperature dependent collinear (XX) Raman spectra of \CGT{} measured in the temperature range of 10 K $ - $ 325 K. T$_{c}$ is indicated by the black dash line.}\label{XX_temp}
\end{figure}

\begin{figure}[!ht]
    \includegraphics[width=\columnwidth]{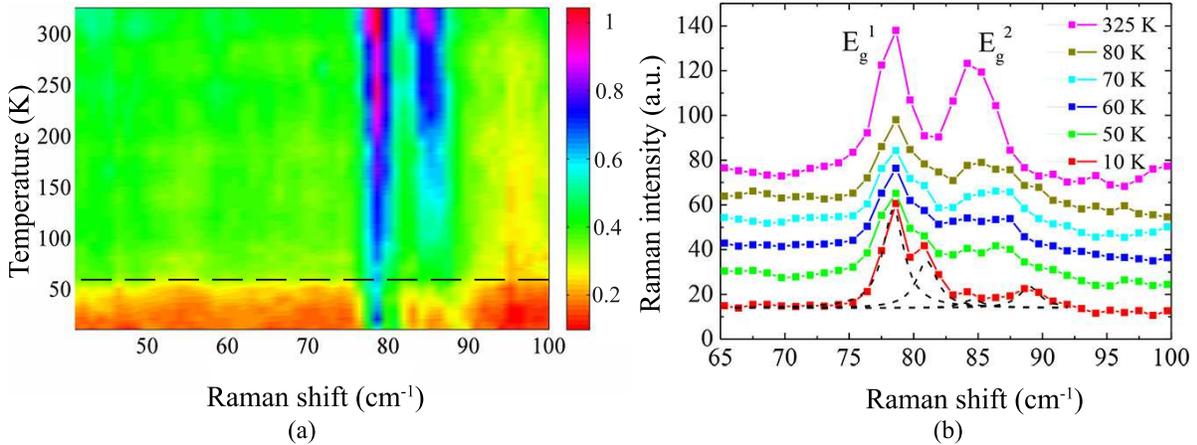}
\caption{(a): Temperature dependent collinear (XX) Raman spectra of \CGT{} for the E$_{g}^{1}$ and E$_{g}^{2}$ modes. T$_{c}$ is indicated by the black dash line. (b): Raw spectra of  E$_{g}^{1}$ and E$_{g}^{2}$ modes. Four Lorentzians (shown in dash line) were used to account for the splitting.}\label{fig:low_energy_colorplot}
\end{figure}

From the Raman spectra of the freshly cleaved \CGT{} sample, we can see that at room temperature there are 7 modes. They center at 78.6 \wavnum, 85.3 \wavnum, 110.8 \wavnum, 136.3 \wavnum, 212.9 \wavnum, 233.9 \wavnum{} and 293.8 \wavnum{} at 270 K. The other three modes might be too weak or out of our spectral range.

To identify the symmetry of these modes, we turn to their polarization dependence (see Fig. \ref{633_532_oldsample_raman}b).
From the Raman tensor (see the supplemental materials), we know that all modes should be visible in the co-linear (XX) geometry and A$_{g}$ modes should vanish in crossed polarized (XY) geometry. To test these predictions we compare the spectra taken at 270 K in XX and XY configurations. As can be seen from Fig. \ref{633_532_oldsample_raman}b, only the two modes located at 136.3 \wavnum{} and  293.8 \wavnum{}  vanish in the XY configuration. Therefore, these two modes are of A$_{g}$ symmetry, and the other five modes are of E$_{g}$ symmetry.

Before proceeding to the temperature dependent Raman studies, it is useful to confirm the quasi-two-dimensional nature of \CGT{}. To achieve this, we exfoliated \CGT{}  on  mica inside an argon filled glovebox to avoid oxidation.  The results are shown in Fig. \ref{fig:exfoliated_CGT}.  We can see  from the optical image (Fig. \ref{fig:exfoliated_CGT}a) that many thin-flake \CGT{} samples can be produced through the mechanical exfoliation method. To verify the thickness, we also performed atomic force microscope (AFM) measurement on two flakes (region 1 and 2). The results are shown in Fig. \ref{fig:exfoliated_CGT}b and  \ref{fig:exfoliated_CGT}c. Both flakes are in nano regime and the region 2 is much thinner than region 1, showing the great promise of preparing 2D \CGT{} samples through this method. To be sure that no dramatic structural changes occur during exfoliation, we also took Raman spectra on the flakes, the results of which are shown in  Fig. \ref{fig:exfoliated_CGT}d. As can be seen from the plot, the Raman spectra of exfoliated \CGT{} are very similar to the bulk, confirming the absence of structural changes. Besides, the Raman intensity of the \CGT{} nanoflakes  increases dramatically as its thickness decreases. This is due to the onset of interference effect and has been observed on many other 2D materials\cite{sandilands2010stability,yoon2009interference,zhao2011fabrication}.

\subsection{Temperature Dependence}

To search for the effects of magnetism (i.e.  magneto-elastic and degeneracy lifting), we measured the Raman spectra well above and below the ferromagnetic transition temperature. The temperature resolution was chosen to be 10 K below 100 K and 20 K above. The full temperature dependence is shown in Fig. \ref{XX_temp}, while we focus on the temperature dependence of the lowest energy E$_{g}$ modes in Fig. \ref{fig:low_energy_colorplot} to search for the lifting of degeneracy due to time reversal symmetry breaking first. Indeed, as the temperature is lowered, additional modes appear in the spectra \splittemperature{} in Fig. \ref{fig:low_energy_colorplot}a. As can be seen more clearly from the Raw spectra in Fig. \ref{fig:low_energy_colorplot}b, the extra feature near the E$_{g}^{1}$ mode and the extremely broad and flat region of the E$_{g}^{2}$ mode appear below \Tc{}. We note that the exact temperature at which this splitting occurs is difficult to determine precisely due to our spectral resolution and the low signal levels of these modes. Nonetheless, the splitting clearly grows as the temperature is lowered and magentic order sets in. At the lowest temperatures we find a 2.9 \wavnum{} splitting for the E$_{g}^{1}$ mode and a 4.5 \wavnum{} splitting for the E$_{g}^{2}$ mode. This confirms our prediction that the lifting of time reversal symmetry leads to splitting of the phonon modes, and suggests significant spin-phonon coupling. Indeed, a similar effect has been observed in numerous three dimensional materials such as MnO\cite{PhysRevB.77.024421}, ZnCr$_{2}$X$_{4}$ (X = O, S, Se)\cite{yamashita2000spin,rudolf2007spin}, and CeCl$_{3}$. In CeCl$_{3}$ the E$_{g}$ symmetry in its point group C$_{4v}$ is also degenerate by time reversal symmetry. For  CeCl$_{3}$ it was found that increasing the magnetic field led to a splitting of two E$_{g}$ modes and a hardening of a second set of E$_{g}$ modes\cite{schaack1977magnetic}. The phonon splitting and energy shifts (discussed in the later section) match well with our observation in \CGT{}.

\begin{figure}[!ht]
   \centering
        \includegraphics[width=0.5\columnwidth]{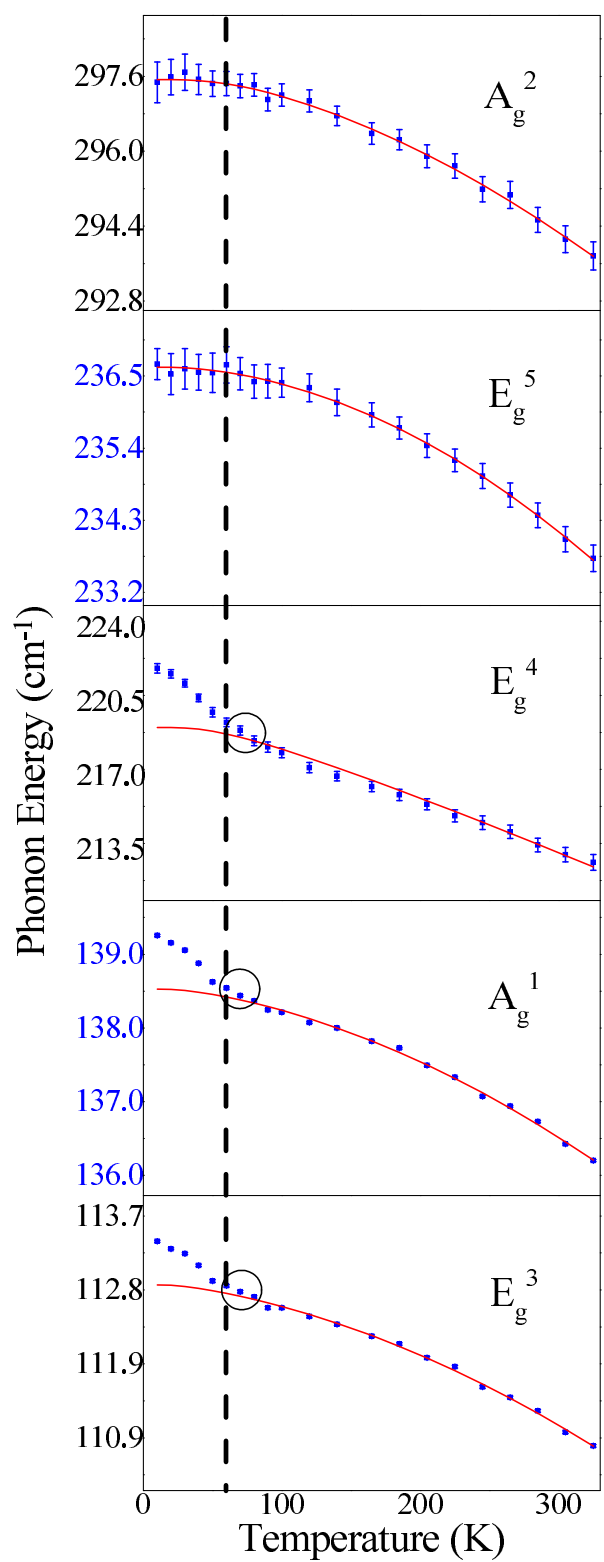}
    \caption{Temperature dependence of phonon frequency shifts. The phonons frequency shifts are shown in blue. The red curves indicate the fit results using the anharmonic model mentioned in the texts. T$_{c}$ is indicated by the dash vertical lines.} \label{CGT_phonon_fits}
\end{figure}

Further evidence of spin-phonon coupling as the origin of the splitting comes for the energy of the modes. The ferromagnetism of \CGT{} originates from the Cr-Te-Cr super-exchange interaction where the Cr octahedra are edge-sharing and the Cr-Te-Cr angle is $91.6^{o}$\cite{CGT_original}. The energy of these two modes are very close to the Te-displacement mode in \CST{}. Thus, it is very likely the E$_{g}^{1}$ and E$_{g}^{2}$ modes involve atomic motions of the Te atoms whose bond strength can be very susceptible to the spin ordering, since the Te atoms mediate the super-exchange between the two Cr atoms.

Before continuing let us consider some alternative explanations for the splitting. For example, structural transitions can also result from magnetic order, however previous X-ray diffraction studies in both the paramagnetic (270 K) and ferromagnetic phases (5 K) found no significant differences in the structure\cite{CGT_original}. Alternatively, the dynamic Jahn-Teller effect can cause phonon splitting\cite{klupp2012dynamic}, but the Cr$^{3+}$ ion is Jahn-Teller inactive in \CGT{}, thus eliminating this possibility as well. One-magnon scattering is also highly unlikely since the Raman tensor of one-magnon scattering is antisymmetric which means the scattering only shows in crossed polarized geometry (XY, XZ and YZ). However, we observed this splitting under XX configuration\cite{fleury1968scattering}.

\begin{figure}[!ht]
   \centering
\includegraphics[width=0.5\columnwidth]{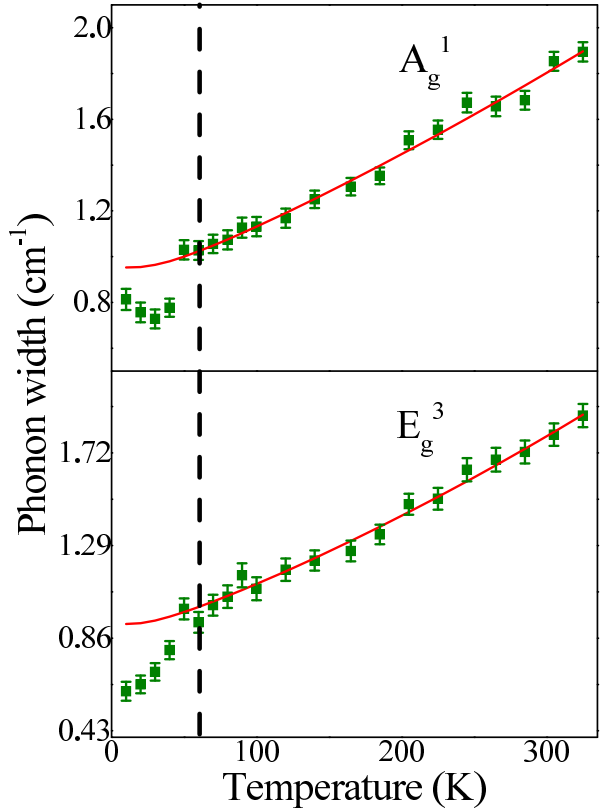}
\caption{Temperature dependence of phonon linewidths (green). The red curves indicate the fit results using equation \ref{eqn:Klemens_model_width} above \Tc{}. The mode located at 110.8 (136.3) \wavnum{} is shown on left (right). T$_{c}$ is indicated by the vertical dash lines.} \label{fig:phononlinewidth}
\end{figure}

Other than the phonon splitting, we also note  a dramatic change in the background Raman scattering at \Tc{} in Fig. \ref{fig:low_energy_colorplot}a.  We believe this is due to  magnetic quasielastic scattering. In a low dimensional magnetic material with spin-phonon coupling, the coupling will induce magnetic energy fluctuation and allow the fluctuations to become observable as a peak centered at 0 \wavnum{} in Raman spectra\cite{reiter1976light,kaplan2006physics}. Typically the peak is difficult to be observed not just due to weak spin-phonon coupling, but since the area under the peak is determined by the magnetic specific heat $C_{m}$ and the width by $D_{t}=C_{m}/\kappa$ where $D_{t}$ is the spin diffusion coefficient, and $\kappa$ is the thermal conductivity. However in low dimensional materials the fluctuations are typically enhanced, increasing the specific heat and lowering the thermal conductivity, making these fluctuations easier to observe in Raman spectra.  This effect has been also observed in many other low dimensional magnetic materials evidenced by the quenching of the scattering amplitude as the temperature drops below \Tc{}\cite{choi2004coexistence,lemmens2003magnetic}.

To further investigate the spin-phonon coupling, we turn our attention to the temperature dependence of the mode frequencies and linewidths. Our focus is on the higher energy modes (E$_{g}^{3}$-A$_{g}^{2}$) as they are easily resolved. To gain more quantitative insights into these modes, we fit the Raman spectra with the Voigt function:
\begin{equation}\label{voigt_function}
  V(x,\sigma,\Omega,\Gamma)=\int^{+\infty}_{-\infty}G(x',\sigma)L(x-x',\Omega,\Gamma)dx'
\end{equation}
which is a convolution of a Gaussian and a Lorentzian\cite{olivero1977empirical}. Here the Gaussian is employed to account for the instrumental resolution and the width $\sigma$ (1.8\wavnum) is determined by the central Rayleigh peak. The Lorentzian represents a phonon mode.
In Fig. \ref{CGT_phonon_fits}, we show the temperature dependence of the extracted phonon energies. All phonon modes soften as the material is heated up. This result is not surprising since the anharmonic phonon-phonon interaction is enhanced at high temperatures and typically leads to a softening of the mode\cite{PhysRevB.29.2051}.  However, for the E$_{g}^{3}$, E$_{g}^{4}$ and A$_{g}^{1}$ modes, their phonon energies change dramatically as the temperature approaches \Tc{}. In fact the temperature dependence is much stronger than we would expect from standard anharmonic interactions.  Especially for the E$_{g}^{4}$ mode, a 2 \wavnum{}  downturn occurs from 10 K to 60 K. This sudden drop of phonon energy upon warming to \Tc{} is a further evidence for the spin-phonon coupling in \CGT{}. Other mechanisms which can induce the shift of phonon energies are of very small probability in this case. For example, an electronic mechanism for the strong phonon energy renormalization is unlikely due to the large electronic gap in \CGT{} (0.202 eV)\cite{Huiwen_doc}. The lattice expansion that explains the anomalous phonon shifts in some magnetic materials\cite{kim1996frequency} is also an unlikely cause. Specifically, the in-plane lattice constant of \CGT{} grows due to the onset of magnetic order\cite{CGT_original}, which should lead to a softening of the modes. However we observe a strong additional hardening of the modes below \Tc{}.

The spin-phonon coupling is also confirmed by the temperature dependence of the phonon linewidths, which are not directly affected by the lattice constants\cite{PhysRevB.29.2051}. In Fig. \ref{fig:phononlinewidth}, we show the temperature dependent phonon linewidths of the E$_{g}^{3}$, and A$_{g}^{1}$ modes due to their larger signal level. We can see the phonon lifetimes are enhanced as the temperature drops below \Tc{}, as the phase space for phonons to scatter into magnetic excitations is dramatically reduced\cite{ulrich2015spin}. This further confirms the spin-phonon coupling.

To further uncover the  spin-phonon interaction, we first remove the effect of the standard anharmonic contributions to the phonon temperature dependence. In a standard anharmonic picture, the temperature dependence of a phonon energy and linewidth is described by:

\begin{eqnarray}
\omega(T)=&\omega_{0}+C(1+2n_{B}(\omega_{0}/2))+   D(1+3n_{B}(\omega_{0}/3)+3n_{B}(\omega_{0}/3)^{2})\\
\Gamma(T)=&\Gamma_{0}+A(1+2n_{B}(\omega_{0}/2))+
B(1+3n_{B}(\omega_{0}/3)+3n_{B}(\omega_{0}/3)^{2})\label{eqn:Klemens_model_width}
\end{eqnarray}
where $\omega_{0}$ is the harmonic phonon energy, $\Gamma_{0}$ is the disorder induced phonon broadening, $n_{B}$ is the Bose-factor, and
$C$ ($A$) and $B$ ($D$) are constants determined by the cubic and quartic anharmonicity respectively. The second term in both equations  results from an optical phonon decaying into two phonons with opposite momenta and a half of the energy of the original mode. The third term describes  the optical phonon decaying into three phonons with a third of the energy of the optical phonon\cite{PhysRevB.28.1928}. The results of fitting the phonon energy and linewidths are shown in red in Fig. \ref{CGT_phonon_fits} and \ref{fig:phononlinewidth} with the resulting parameters listed in in table \ref{table:Anharmonic_fit_data}. We can see that the temperature dependent frequencies of the two highest energy modes E$_{g}^{5}$, A$_{g}^{2}$ follow the anharmonic prediction very well throughout the entire range. However, for the other three modes, there is a clear deviation from the anharmonic prediction below \Tc{} confirming the existence of spin-phonon coupling. Moreover, we notice that for the E$_{g}^{3}$, A$_{g}^{1}$ and E$_{g}^{4}$ modes, the phonon energies start to deviate from the anharmonic prediction even above \Tc{} (circled in Fig. \ref{CGT_phonon_fits}). This is probably due to the short-ranged two-dimensional magnetic correlations that persist to temperatures above \Tc{}. Indeed finite magnetic moments\cite{Huiwen_doc} and magneto-striction\cite{CGT_original} were observed in \CGT{} above \Tc{}.
\begin{table}[!ht]
  \caption{Anharmonic interaction parameters. The unit is in \wavnum.}\label{table:Anharmonic_fit_data}
  \centering
 \begin{tabular}{c c c c c c c c c c c}
 \hline\hline
Mode & $\omega_{0}$ &Error &C &Error& D &Error &A &Error& B& Error\\ \hline
E$_{g}^{3}$ & 113.0 &    0.1 &   -0.09 &    0.03 &   -0.013 &    0.002 & 0.10 &   0.03   &0.003 &  0.002\\
A$_{g}^{1}$ & 138.7 &    0.1 &   -0.12 &    0.04 &   -0.024 &    0.003 & 0.13  & 0.03 & 0.003  & 0.003\\
E$_{g}^{4}$ & 220.9 &    0.4 &   -1.9 &    0.1 &    -0.02 &    0.01 &-- &--&--&--\\
E$_{g}^{5}$ & 236.7 &    0.1 &   -0.01 &    0.07 &   -0.12 &    0.01 &-- &--&--&--\\
A$_{g}^{2}$ & 298.1 &    0.3 &   -0.4 &    0.3 &   -0.20 &    0.04 &-- &--&--&--\\ \hline
\end{tabular}
\end{table}

\section{Discussion}
The spin-phonon coupling in 3d-electron systems usually results from the modulation of the electron hopping amplitude by ionic motion, leading to a change in the exchange integral $J$. In the unit cell of \CGT{} there are two in-equivalent magnetic ions (Cr atoms), therefore the spin-phonon coupling Hamiltonian to the lowest order can be written as\cite{woods2001magnon,PhysRev.127.432},

\begin{equation}
\spinphononcoupling
\end{equation}
where $\mathbf{S}$ is a spin operator, $u$ stands for the ionic displacement of atoms on the exchange path, the index ($i$) runs through the lattice, $\delta$ is the index of its adjacent sites, and the subscripts $a$ and $b$ indicate the in-equivalent Cr atoms in the unit cell. The strength of the coupling to a specific mode depends on how the atomic motion associated with that mode, modulates the exchange coupling. This in turn results from the detailed hybridization and/or overlap of orbitals on different lattice sites. Thus, some phonon modes do not show the coupling effect regardless of their symmetry.

To extract  the spin-phonon coupling coefficients, we use a simplified version of equation \ref{equ:hamitionian_sp}\cite{fennie2006magnetically,lockwood1988spin},
\begin{equation}
\omega\approx\omega_{0}^{ph}+\lambda{}<\mathbf{S}_{i}^{a}{}\cdot\mathbf{S}_{i+\delta}^{b}>\label{spin_phonon_coupling_equation}
\end{equation}
where $\omega$ is the frequency of the phonon mode, $\omega_{0}^{ph}$ is the phonon energy free of the spin-phonon interaction, $<\mathbf{S}_{i}^{a}\cdot\mathbf{S}_{i+\delta}^{b}>$ denotes a statistical average for adjacent spins, and $\lambda$ represents the strength  of the spin-phonon interaction which is proportional to $\frac{\partial{}J}{\partial{}u}u$. The saturated magnetization value of \CGT{} reaches 3$\mu$B per Cr atom at 10 K, consistent
with the expectation for a high spin configuration state of Cr$^{3+}$\cite{Huiwen_doc}. Therefore, $<\mathbf{S}_{i}^{a}{}\cdot\mathbf{S}_{i+\delta}^{b}> \approx 9/4$ for Cr$^{3+}$ at 10 K and the spin-phonon coupling constants can be estimated using equation \ref{spin_phonon_coupling_equation}.  The calculated results are given in table \ref{spin_phonon_table}.

Compared to the geometrically (\CCO, \ZCO) or bond frustrated (\ZCS) chromium spinels the coupling constants are smaller in \CGT{}\cite{rudolf2007spin}. This is probably not surprising, because  in the spin frustrated materials, the spin-phonon couplings are typically very  strong\cite{rudolf2007spin}.  On the other hand,  in comparison with the cousin compound \CST{} where the coupling constants were obtained for the phonon modes at 90.5 \wavnum{} ($\lambda$=0.1) and 369.3 \wavnum{} ($\lambda$=-0.2)\cite{casto2015strong}, the coupling constants in \CGT{} is larger.   

\begin{table}[!ht]
  \caption{Spin-phonon interaction parameters at 10 K. The unit is in \wavnum.}
  \centering
  \begin{tabular}{c c c c c c}
  \hline\hline
  Mode & $\omega$ & $\omega_{0}^{ph}$ & $\lambda$\\[0.5ex]
  \hline
   E$_{g}^{3}$ & 113.4 & 112.9 & 0.24 \\
   A$_{g}^{1}$ & 139.3 & 138.5 & 0.32\\
   E$_{g}^{4}$ & 221.7 & 219.0 & 1.2 \\
  \hline
\end{tabular}\label{spin_phonon_table}
\end{table}

\section{\label{sec:exp}Conclusion}
In summary, we have demonstrated spin-phonon coupling in a potential 2D atomic crystal for the first time. In particular we studied the polarized temperature dependent Raman spectra of \CGT{}. The two lowest energy modes of E$_{g}$ symmetry split below \Tc{}, which is ascribed to the time reversal symmetry breaking by the spin ordering.  The temperature dependence of the five modes at higher energies were studied in detail revealing additional evidence for spin-phonon coupling. Among the five modes, three modes show significant renormalization of the phonon lifetime and frequency due to the onset of magnetic order. Interestingly, this effect appears to emerge above \Tc{}, consistent with other evidence for the onset of magnetic correlations at higher temperatures. Besides,  magnetic quasielastic scattering was also observed in \CGT{}, which is consistent with the spin-phonon coupling effect. Our results  also show the possibility to study magnetism in exfoliated 2D ferromagnetic \CGT{} from the perspective of the phonon modes and magnetic quasielastic scattering using micro-Raman scattering.

\section{Acknowledgements}
We are grateful for numerous discussions with Y. J. Kim and H. Y. Kee  at University of Toronto. Work at University of Toronto was supported by NSERC, CFI, and ORF and K.S.B. acknowledges support from the National Science Foundation (Grant No. DMR-1410846). The crystal growth at Princeton University was supported by the NSF MRSEC Program, grant number NSF-DMR-1005438.

\section{References}
\providecommand{\newblock}{}

\end{document}